\documentclass[preprint,prd,aps,a4paper,10pt,titlepage]{revtex4-1} 
\usepackage{color}
\usepackage{graphicx}
\usepackage{amsmath}

\newcommand{\eps}{\varepsilon}

\begin{document}

\vglue 5mm

\title{Theoretical Analysis of the $p \bar{p}\rightarrow \pi^0  e^+  e^-$ Process within a Regge Framework}

\author{J. Guttmann}

\author{M. Vanderhaeghen}
\affiliation{Institut f\"ur Kernphysik, Johannes Gutenberg-Universit\"at, Mainz D-55099, Germany}

\date{\today}
\begin{abstract}
We study the annihilation process $p\bar{p} \to \pi^0 e^+ e^-$ within a Regge framework, as a means to provide constraints on timelike nucleon form factors. We present results for the $e^+e^-$ angular distributions and the differential cross sections in kinematics which will be accessible by PANDA@FAIR. To check the consistency of the model, we first test the approach on the process of real photon production, $\bar{p} p \to \pi^0 \gamma$, where data in the energy range of $2.911$ GeV$\leq\sqrt{s} \leq 3.686$ GeV exist. We find that a Regge pole model is able to well reproduce the available data. The analysis is then extended to a timelike virtual photon in the final state.
\end{abstract}
%
\maketitle


\section{Introduction}
The structure of the nucleon in terms of the electromagnetic form factors can be probed by the electromagnetic interaction. In the spacelike region of negative momentum-transfer $q^2$ the electric and magnetic form factors, $G_E$ and $G_M$, can be investigated in elastic electron-proton scattering. The spacelike form factors, which provide information on the electromagnetic spatial distributions in the nucleon, have been studied extensively, using both unpolarized measurements as well as polarization experiments, for reviews see e.g. Refs.~\cite{HydeWright:2004gh,Arrington:2006zm,Perdrisat:2006hj}. The corresponding crossed processes $p\bar{p} \rightarrow e^+e^-$ or $e^+e^- \rightarrow p\bar{p}$ allow to access the form factors in the timelike region ($q^2>0$), starting from the threshold $q^2=4m_N^2$, where $m_N$ is the nucleon mass.
A few measurements of the latter processes, mainly of the total cross section, exist. Only two experiments, performed at BaBar \cite{Aubert:2005cb} and LEAR \cite{Bardin:1994am}, attempted an individual extraction of the timelike electric and magnetic form factors.
New measurements of the timelike form factors are planned at BES-III and PANDA@FAIR~\cite{Lutz:2009ff} and will explore the at present still largely uncharted timelike region in much greater detail. However, the timelike region below the threshold $0<q^2<4m_N^2$, which is denoted as the unphysical region, is not accessible by annihilation processes as $p\bar{p} \rightarrow e^+e^-$ or $e^+e^- \rightarrow p\bar{p}$. Nevertheless, to obtain a consistent picture of the nucleon, knowledge of the electromagnetic form factors over the full range of $q^2$ is needed.

Therefore, in Refs.~\cite{Dubnickova:1995ns, Adamuscin:2007iv} the annihilation process, where in addition a neutral pion is produced, $p\bar{p} \rightarrow \pi^0 e^+e^-$ has been studied. Since the outgoing pion takes a part of the energy of the process, the production of a lepton pair with an invariant mass below the $(p+\bar{p})$-annihilation threshold is possible and thus this reaction can be used to study the electromagnetic form factors in the unphysical region. Moreover this reaction provides the possibility to access the relative phases of $G_E$ and $G_M$, in contrast to cross section measurements of $\bar{p}p\rightarrow e^+ e^-$, where only the modulus of the form factors can be extracted. Feasibility studies of the process with PANDA@FAIR have been performed in \cite{Boucher2011}.
However, a study of the timelike form factors from the $p\bar{p} \to \pi^0e^+e^-$ process requires a theory in order to extrapolate to the nucleon pole.

In Ref.~\cite{Adamuscin:2007iv}, the process has been calculated using a Born diagram model, in which the reaction is described through the exchange of a single nucleon, the Feynman diagrams are shown in Fig.~\ref{fi:bornmodel}. A formula for the differential cross section has been presented, where the dependence on the lepton kinematics has been removed by integration over the full phase space of the leptonic subprocess. The calculation has been extended in \cite{Gakh:2012uk} for an exclusive experimental setup including the investigation of polarization observables. For a high invariant mass lepton pair the process has been studied in \cite{Lansberg:2007se,Lansberg:2012ha} using the concept of transition distribution amplitudes.

In this paper, we study the process  $p\bar{p} \rightarrow \pi^0 e^+e^-$ using Regge theory \cite{Collins:1977jy, Storrow:1986zw}, and describe the process at high energy within a Regge pole model. This phenomenological approach has been successfully applied to electroproduction and photoproduction of pions and kaons, see e.g. \cite{Guidal:1997hy,Vanderhaeghen:1997ts}. In particular, it has been widely applied in order to extract $\pi^+$ and $K^+$ electromagnetic form factors from the $\pi^+$ and $K^+$ electroproduction process \cite{Blok:2008jy,Coman:2009jk}. To check the consistency of this model, we first test the approach on the process of real photon production, $\bar{p} p \to \pi^0 \gamma$, which has been measured at Fermilab \cite{Armstrong:1997gv}, and compare the results of our calculation with the existing data. We find that the Regge pole model is able to describe the results of the measurements, whereas the Born diagram model cannot reproduce the available data.

After specifying the Regge model for $\bar{p} p \to \pi^0 \gamma$, we extend the approach to the annihilation process $\bar{p} p \to \pi^0 e^+ e^-$ and investigate the cross section with regard to the determination of the form factors in the unphysical region within the kinematic range, which will be accessible by PANDA@FAIR. 
Furthermore a general expression of the cross section including the dependence on the full lepton pair kinematics is presented.

The paper is organized as follows: in Section \ref{sec:pi0gamma}, we analyze the process $p\bar{p} \to \pi^0 \gamma$. We introduce the Regge pole model and compare our results to the existing data. In Section \ref{sec:pi0ll} the approach is applied to $p\bar{p} \to \pi^0 e^+ e^-$. The cross section in terms of the angular distribution of the produced lepton pair is given and the angular dependence of the cross section for several kinematic setups is presented within the Regge model. We summarize our findings in Section \ref{conclusions}.



\section{The $\bar{p}p \rightarrow \pi^0 \gamma$ Process within a Regge Framework}
\label{sec:pi0gamma}
Before describing the theoretical model for $p \bar{p} \rightarrow \pi^0 e^+ e^-$ process, for which at present no data exist, we firstly consider the reaction
\begin{equation}
\bar{p}(p_1) +  p(p_2) \rightarrow \pi^0(q_\pi) + \gamma(q),
\end{equation}
which has been measured at Fermilab \cite{Armstrong:1997gv}, and will test our model predictions with the results of this experiment. Data of the angular dependence of the differential cross section $d\sigma/d\cos\theta_\pi$ is available in the center-of-mass (c.m.) energy range of $2.911$ GeV$\leq\sqrt{s} \leq 3.686$ GeV.

\begin{figure}
\begin{center}
 \includegraphics[width=5.0cm]{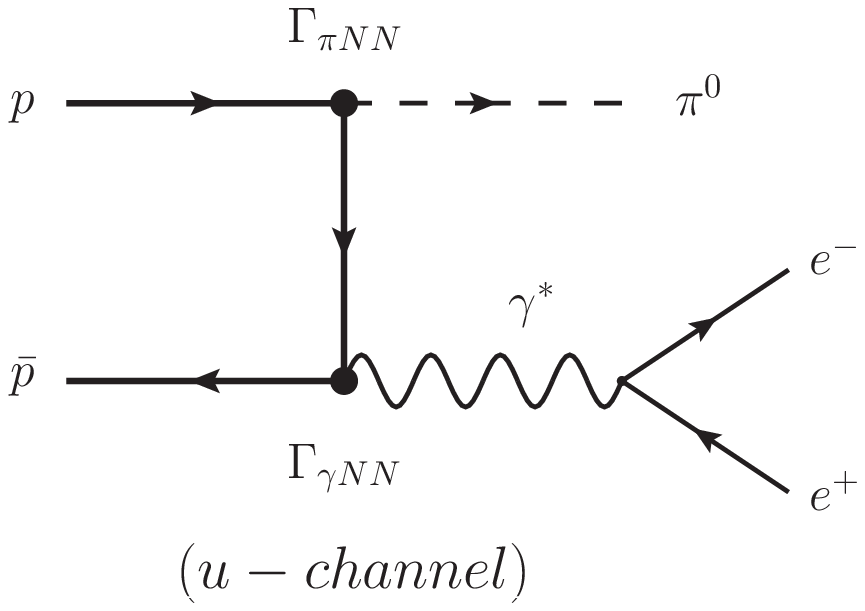}
 \includegraphics[width=5.0cm]{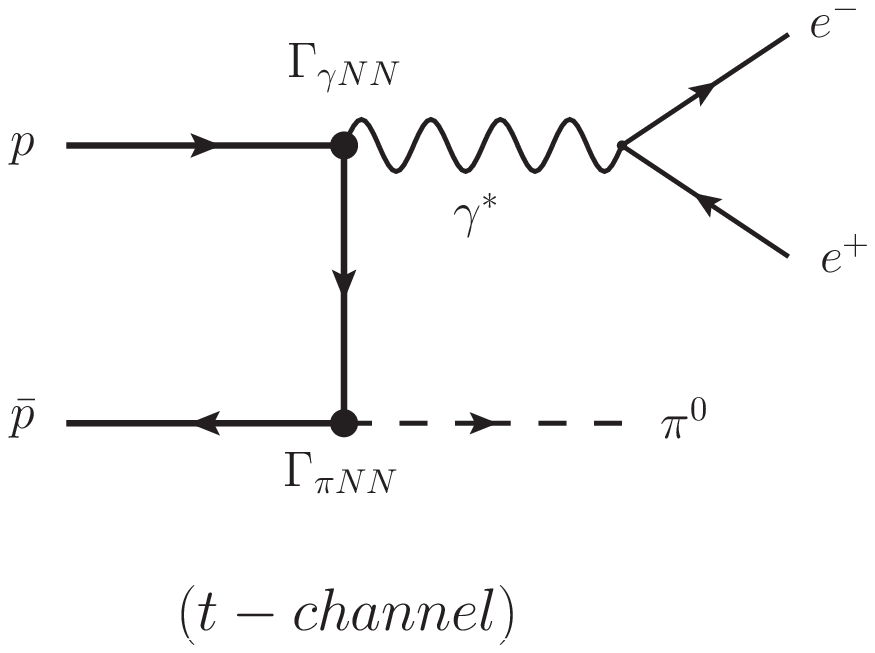}
\end{center}
\caption{Born model for $p \bar{p} \rightarrow \pi^0 e^+ e^-$ described by a single nucleon exchange in the $u$-channel and $t$-channel diagrams. \label{fi:bornmodel}}
\end{figure}

The annihilation process can be described by two kinematic invariants, e.g. two of the Mandelstam variables, defined as:
\begin{equation}
 s=(p_1+p_2)^2, \quad u=(p_2-q_\pi)^2, \quad t=(p_1-q_\pi)^2.
\end{equation}

We start with the model considered in Ref.~\cite{Adamuscin:2007iv}, where the process $\bar{p}p \to \pi^0 e^+ e$ has been described within a Born diagram model through the exchange of a single nucleon. The two corresponding Feynman diagrams, which are shown in Fig.~\ref{fi:bornmodel}, are given by a $u$-channel and $t$-channel nucleon exchange. The amplitude of the diagrams can be written as
\begin{align}
\begin{split}
 \mathcal{T}_u =\frac{1}{q^2}\,\mathcal{L}_\mu \cdot \mathcal{M}^\mu_{\pi^0\gamma^*,\ u} = & \frac{1}{q^2}\, \mathcal{L}_\mu \cdot \bar{v}(p_1)\,\Gamma_{\gamma N N}^\mu(q)\left(\frac{\gamma\cdot(p_2-q_\pi)+m_N}{u-m_N^2}\right)\Gamma_{\pi NN}(q_\pi)\,u(p_2), \\
 \mathcal{T}_t = \frac{1}{q^2}\,\mathcal{L}_\mu \cdot \mathcal{M}^\mu_{\pi^0\gamma^*,\ t} = & \frac{1}{q^2}\,\mathcal{L}_\mu \cdot \bar{v}(p_1)\,\Gamma_{\pi N N}(q_\pi)\left(\frac{\gamma\cdot(q_\pi-p_1)+m_N}{t-m_N^2}\right)\Gamma_{\gamma NN}^\mu(q)\,u(p_2),
\end{split}
\end{align}
where $\Gamma_{\gamma NN}$ ($\Gamma_{\pi NN}$) is the parametrization of the $\gamma N N^*$ ($\pi N N^*$) vertices and $\mathcal{L}_\mu$ describes the leptonic subprocess $\gamma^* \rightarrow e^+ e^-$.

Within this approach off-shell effects of the exchanged nucleons were neglected, hence the $\gamma NN^*$-vertices are parametrized by the on-shell proton electromagnetic form factors, in terms of the Dirac and Pauli form factors $F_1$ and $F_2$ given by
\begin{equation}
\Gamma_{\gamma N N}^\mu(q) = e\left[F_1(q^2)\gamma^\mu - \frac{i}{2m_N}F_2(q^2)\sigma^{\mu\nu}q_\nu \right].
\end{equation}
Analogously to the findings of photoproduction and electroproduction of pions at high energies \cite{Guidal:1997hy,Vanderhaeghen:1997ts}, we consider a $\pi NN$ coupling of the pseudoscalar type, as
\begin{equation}
 \Gamma_{\pi NN}(q_\pi) = g_{\pi NN}(m_\pi^2) \gamma_5,
\label{eq:piNN}
\end{equation}
where  $g_{\pi NN}(m_\pi^2)$ is the pion-nucleon coupling constant.

In case of real photon production $\bar{p}p \rightarrow \pi^0 \gamma$, the amplitudes within the Born diagram model are obtained as
\begin{align}
\begin{split}
 \mathcal{T}_u =\eps^*_\mu \cdot \mathcal{M}_{\pi^0\gamma,\ u}^{\mu} = & \eps^*_\mu(q,\lambda_\gamma) \cdot \bar{v}\,(p_1)\Gamma_{\gamma N N}^\mu(q)\left(\frac{\gamma\cdot(p_2-q_\pi)+m_N}{u-m_N^2}\right)\Gamma_{\pi NN}(q_\pi)\,u(p_2), \\
 \mathcal{T}_t = \eps^*_\mu \cdot \mathcal{M}^\mu_{\pi^0\gamma,\ t} = & \eps^*_\mu(q, \lambda_\gamma) \cdot \bar{v}(p_1)\,\Gamma_{\pi N N}(q_\pi)\left(\frac{\gamma\cdot(q_\pi-p_1)+m_N}{t-m_N^2}\right)\Gamma_{\gamma NN}^\mu(q)\,u(p_2),
\end{split}
\end{align}
where $\eps(q,\lambda_\gamma)$ is the real photon polarization vector and the form factors $F_1(q^2=0)=1$ and $F_2(q^2=0)=\kappa_P=1.79$ have been used in the description of $\Gamma_{\gamma NN}$.

The differential cross section $d\sigma/dt$ for the process $p\bar{p}\to \pi^0\gamma$ can be expressed as:
\begin{align}
\frac{d\sigma}{dt} =  \frac{1}{16\pi s(s-4m_N^2)}(-g_{\mu\nu})\mathcal{H}^{\mu\nu}_{\pi^0\gamma},
\end{align}
where $\mathcal{H}^{\mu\nu}_{\pi^0\gamma}$ is the hadronic tensor, given by
\begin{equation}
 \mathcal{H}^{\mu\nu}_{\pi^0\gamma}=\frac{1}{4}\sum_{\mathrm{spins}}\big|\mathcal{M}^\mu_{\pi^0\gamma,\,u}+\mathcal{M}^\mu_{\pi^0\gamma,\,t}\big|^2.
\end{equation}

Using the Born diagram model for $\bar{p}p \rightarrow \pi^0 \gamma$, we are not able to reproduce the results of the E760 experiment. The obtained cross section $d\sigma/d\cos\theta_\pi$ is about 4 to 5 orders of magnitude larger than the data, depending on the value of the c.m. energy $\sqrt{s}$. Simple fixes by introducing strong suppressions through off-shell form factors do not lead to a correct energy dependence of the cross sections. One can assume, that this model is not suitable to describe this process and thus the process $\bar{p} p \to \pi^0 e^+ e^-$ as well.

Therefore we consider a Regge pole description, which is based on the exchange of dominant baryon Regge trajectories in the $u$-channel and $t$-channel. This approach is valid in the kinematic region at forward and backward angles, $s \gg |t|$ and $s \gg |u|$. In the kinematical region $s\sim-t\sim-u$ the reaction has been investigated within the framework of generalized distribution amplitudes \cite{Kroll:2005ni}. The dominant trajectories for the process $\bar{p}p\rightarrow \pi^0\gamma$ are the nucleon ($N$) trajectory and $\Delta$-trajectory associated with the $\Delta(1232)$-resonance. The amplitude for Regge trajectory exchange can be obtained from the Born diagram by replacing the usual Feynman propagator of the single exchanged particle by the so-called Regge propagator, while leaving the Feynman structure, giving rise to the residue of the Regge pole, unchanged.

In case of an exchanged nucleon, the pole-like Feynman propagators of the $u$-channel and $t$-channel, given by $1/(u-m_N^2)$ and $1/(t-m_N^2)$, are replaced in the following way
\begin{align}
\begin{split}
  \frac{1}{u-m_N^2} \ \Rightarrow & \ D^{\mathrm{Regge}}_N(u,s) =  \frac{s^{\alpha_N(u)-\frac{1}{2}}}{\Gamma\left[\alpha_N(u)+\frac{1}{2}\right]}\pi\alpha_N^\prime\frac{\mathrm{e}^{-i\pi\left(\alpha_N(u)+\frac{1}{2}\right)}}{\sin\pi\left(\alpha_N(u)+\frac{1}{2}\right)} \\[1mm]
  \frac{1}{t-m_N^2} \  \Rightarrow & \ D^{\mathrm{Regge}}_N(t,s) =  \frac{s^{\alpha_N(t)-\frac{1}{2}}}{\Gamma\left[\alpha_N(t)+\frac{1}{2}\right]}\pi\alpha_N^\prime\frac{\mathrm{e}^{-i\pi\left(\alpha_N(t)+\frac{1}{2}\right)}}{\sin\pi\left(\alpha_N(t)+\frac{1}{2}\right)}
\end{split}
\end{align}
where the nucleon trajectory $\alpha_N$ is of the form $\alpha_N(u) = \frac{1}{2} + \alpha^\prime_N(u-m_N^2)$ with $\alpha^\prime_N=0.97$ GeV$^{-2}$.

\begin{figure}
\begin{center}
 \includegraphics[width=7.8cm]{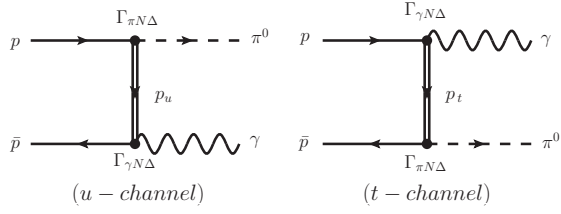}
\end{center}
\caption{Born model for $p + \bar{p} \rightarrow \pi^0 + \gamma$ described by the exchange of $\Delta(1232)$. \label{fi:delta}}
\end{figure}

Besides including the nucleon Regge propagators, we consider the exchange of the $\Delta$-trajectory. Starting from the Feynman diagrams in a Born model, illustrated in Fig.~\ref{fi:delta}, the amplitudes of the exchange of the $\Delta$(1232)-resonance can be expressed by
\begin{align}
\begin{split}
\mathcal{M}^{\Delta,\,\mu}_{\pi^0\gamma,u} =&  \bar{v}(p_1)\Gamma^\alpha_{\gamma N\Delta}\frac{-i}{u-m_\Delta^2}(\gamma\cdot p_u + m_\Delta)  \\[1mm]
& \times \left\{g_{\alpha\beta} -\frac{1}{3}\gamma_\alpha\gamma_\beta -\frac{\gamma_\alpha p_{u,\beta} -\gamma_\beta p_{u,\alpha}}{3m_\Delta} -\frac{2p_{u ,\alpha}\,p_{u,\beta}}{3m_\Delta^2} \right\}\Gamma^\beta_{\pi N \Delta} u(p_2),  \\[1mm]
\mathcal{M}^{\Delta,\,\mu}_{\pi^0\gamma,t} =& \bar{v}(p_1)\Gamma^\alpha_{\pi N\Delta}\frac{-i}{t-m_\Delta^2}(\gamma\cdot p_t + m_\Delta) \\[1mm]
& \times \left\{g_{\alpha\beta} -\frac{1}{3}\gamma_\alpha\gamma_\beta -\frac{\gamma_\alpha p_{t,\beta} -\gamma_\beta p_{t,\alpha}}{3m_\Delta} -\frac{2p_{t ,\alpha}\,p_{t,\beta}}{3m_\Delta^2} \right\}\Gamma^\beta_{ \gamma N \Delta} u(p_2),
\end{split}
\label{eq:deltaam}
\end{align}
with $p_u=p_2-q_\pi$, and $p_t = q_\pi-p_1$. $\Gamma_{\gamma N \Delta}$ and $\Gamma_{\pi N \Delta}$ are describing the $\gamma N \Delta$-vertices and  $\pi N \Delta$-vertices, respectively, and can be parametrized through
\begin{align}
 \Gamma_{\gamma N \Delta}^\alpha =& i\sqrt{\frac{2}{3}}\frac{3e(m_\Delta + m_N)}{2m_N((m_\Delta + m_N)^2 -q^2)}g_M(q^2)\varepsilon^{\alpha \mu \rho \sigma}\, p_{\Delta,\rho}\,q_{\sigma}, \\[1mm]
\Gamma_{\pi N \Delta}^\alpha =& -\frac{h_A}{2f_\pi m_\Delta}\gamma^{\alpha\mu\nu}q_{\pi,\mu}\,p_{\Delta, \nu} T_a^\dagger,
\label{eq:deltav}
\end{align}
where $p_\Delta$ is the 4-momentum of the intermediate $\Delta$-state. The operator $T_a^\dagger$ is the isospin $1/2 \rightarrow 3/2$ transition operator, $f_\pi$ denotes the pion decay constant and $h_A \simeq 2.85$ the $\pi N\Delta$ coupling constant. In  Eq.~(\ref{eq:deltav}) the $\gamma N\Delta$ vertex is parametrized by the form factor $g_M(q^2)$, the magnetic dipole form factor representing the strength of the magnetic dipole $N\rightarrow \Delta$ transition. The electric quadrupole and Coulomb quadrupole terms have been neglected in our calculation since their contributions have been found to be of order of a few \% \cite{Tiator:2011pw}. We will use as $\gamma N \Delta$ coupling strength $g_M(0) = 3.02$.

The Feynman propagators in Eq.~(\ref{eq:deltaam}) are then replaced by the Regge propagators:
\begin{align}
\begin{split}
  \frac{1}{u-m_\Delta^2}  \ \Rightarrow & \ D^{\mathrm{Regge}}_\Delta(u,s) =  \frac{s^{\alpha_\Delta(u)-\frac{3}{2}}}{\Gamma\left[\alpha_\Delta(u)+\frac{1}{2}\right]}\pi\alpha_\Delta^\prime\frac{\mathrm{e}^{-i\pi\left(\alpha_\Delta(u)-\frac{1}{2}\right)}}{\sin\pi\left(\alpha_\Delta(u)-\frac{1}{2}\right)}, \\[1mm]
  \frac{1}{t-m_\Delta^2}  \  \Rightarrow & \ D^{\mathrm{Regge}}_\Delta(t,s) =  \frac{s^{\alpha_\Delta(t)-\frac{3}{2}}}{\Gamma\left[\alpha_\Delta(t)+\frac{1}{2}\right]}\pi\alpha_\Delta^\prime\frac{\mathrm{e}^{-i\pi\left(\alpha_\Delta(t)-\frac{1}{2}\right)}}{\sin\pi\left(\alpha_\Delta(t)-\frac{1}{2}\right)},
\end{split}
\end{align}
where the $\Delta$-Regge trajectory is of the form $\alpha_\Delta(u)=\frac{3}{2} +\alpha_\Delta^\prime(u-m_\Delta^2)$, with $\alpha_\Delta^\prime = 0.9$ GeV$^{-2}$.

We reggeize the amplitude of the process in the following way, which ensures gauge invariance of the Regge model amplitudes:
\begin{align}
 \mathcal{M}^N_{\pi^0\gamma,t} &=  D^{\mathrm{Regge}}_N(t,s)(t-m_N^2)\ [\mathcal{M}_u + \mathcal{M}_t], \label{eq:ampMt}\\
 \mathcal{M}^N_{\pi^0\gamma,u} &=  D^{\mathrm{Regge}}_N(u,s)(u-m_N^2)\,[\mathcal{M}_u + \mathcal{M}_t],  \label{eq:ampMu}
\end{align}
and analogous expressions for the $\Delta$-exchange diagrams. Note that the Regge approach implies $s\gg|t|$, $s\gg |u|$, so that both forward and backward regions are kinematically separated. In the kinematic region $s\gg |t|$ the Regge amplitude of Eq.~(\ref{eq:ampMt}) is dominating, whereas in region of $s\gg |u|$ the $u$-channel Regge amplitude (Eq.~(\ref{eq:ampMu})) is the dominant one. Only in this limits there is no double counting in this procedure. In the intermediate angular region one is outside the range of the validity of a Regge approach.
In the limits $u \rightarrow m_N^2$ or $t \rightarrow m_N^2$, the results of the Born diagram model are recovered, since one obtains $D_N^{\mathrm{Regge}}(u,s)\cdot (u-m_N^2) \rightarrow 1$ and  $D_N^{\mathrm{Regge}}(t,s)\cdot (t-m_N^2) \rightarrow 1$ in this limits, respectively.

We next discuss the inclusion of the $\Delta$-exchange Regge trajectories. As for the $\Delta$ we are further away from the pole position than in the nucleon case, the description of the residues of the Regge poles through their on-shell couplings can be expected to be modified. We allow for such a reduction of the coupling strengths of the $\Delta$ Regge pole residue, leading to the amplitude
\begin{equation}
 \mathcal{H}_{\pi^0\gamma} = \frac{1}{4}\sum_{\mathrm{spins}}\left|\mathcal{M}^N_{\pi^0\gamma}+\mathcal{F}\cdot\mathcal{M}^\Delta_{\pi^0\gamma}\right|^2,
\label{eq:ampF}
\end{equation}
where the parameter $\mathcal{F}$ will be obtained by a fit to the data.

\begin{figure*}
\includegraphics[width=7.8cm]{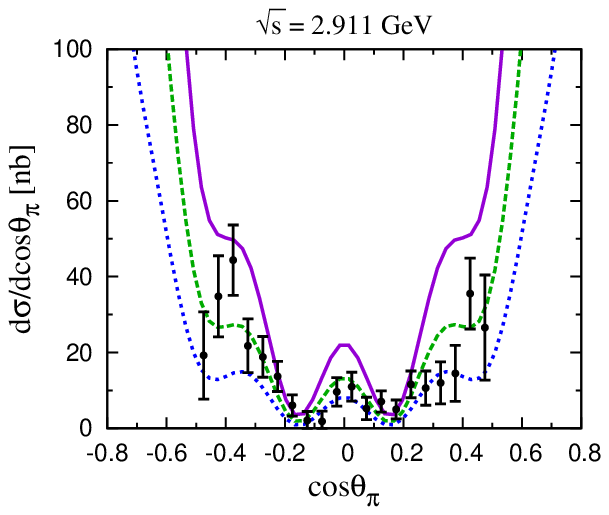}
\includegraphics[width=7.8cm]{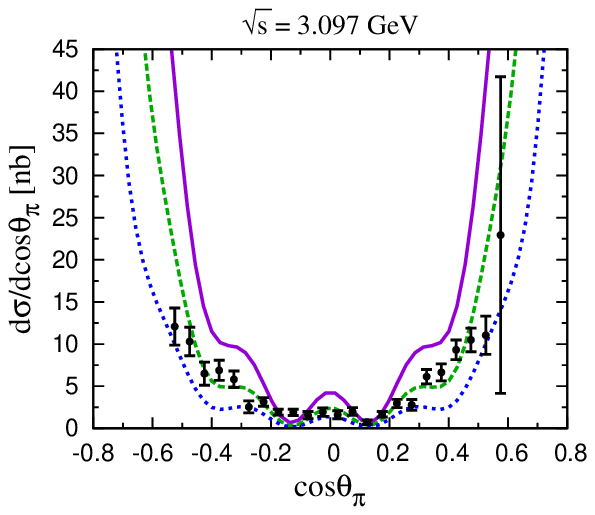}\\
\includegraphics[width=7.8cm]{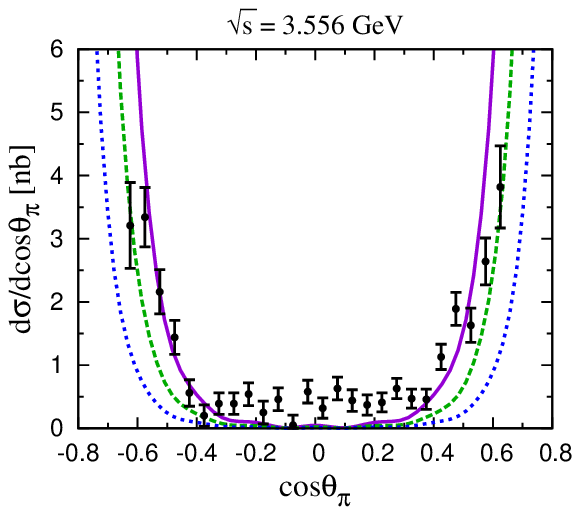}
\includegraphics[width=7.8cm]{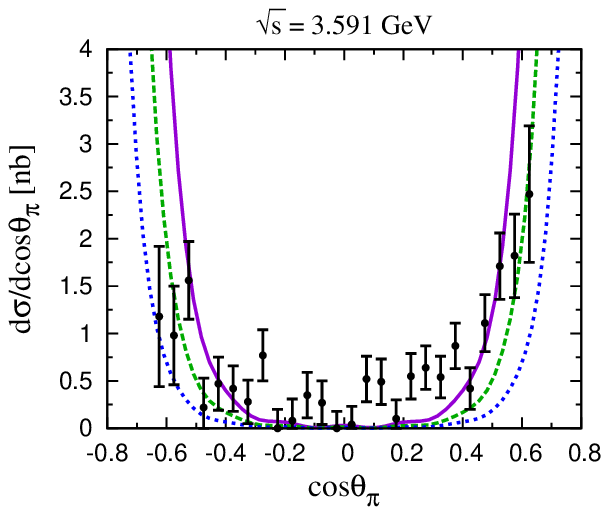} \\
\includegraphics[width=7.8cm]{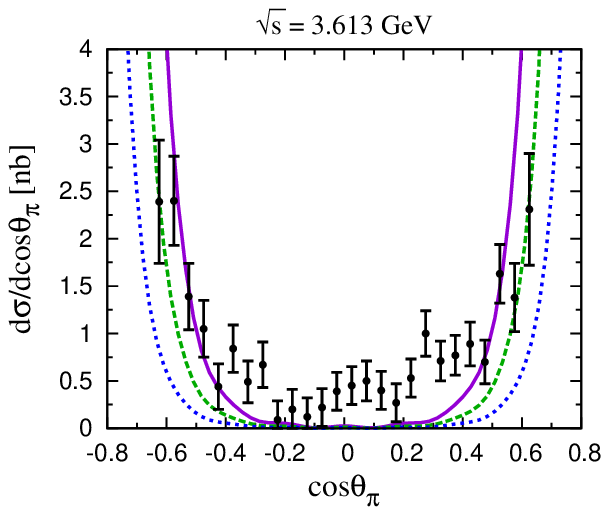}
\includegraphics[width=7.8cm]{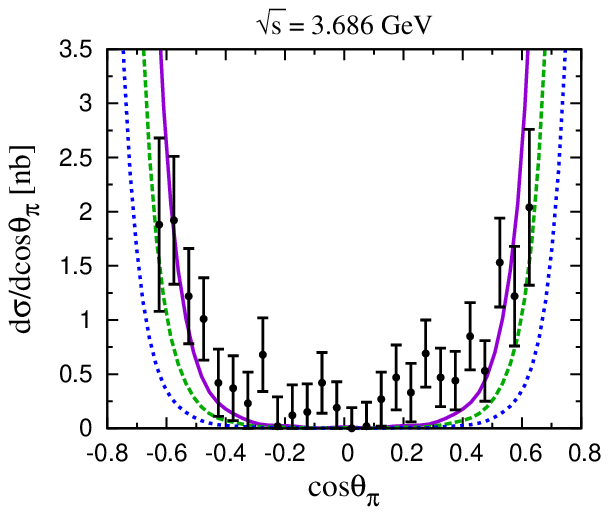}
\caption{Differential cross section $d\sigma/d\cos\theta_\pi$ of $\bar{p}p \rightarrow \pi^0\gamma$ for different c.m. energies $\sqrt{s}$; blue (dotted) curve: $N$-trajectory contribution; purple (solid) curve: cross section including $(N+\Delta)$ Regge trajectory exchange; green (dashed) curve: $(N+\Delta)$ contribution including a reduction of the $\Delta$ pole residue ($\mathcal{F} \approx 0.5$) according to Eq.~(\ref{eq:ampF}). The data are from Ref.~\cite{Armstrong:1997gv}. \label{fi:pi0gamma_ND}}
\end{figure*}

In Fig.~\ref{fi:pi0gamma_ND}, results for $d\sigma/d\cos\theta_\pi$ including $N$-trajectory exchange and $(N+\Delta)$-trajectories exchange are presented as well as results using the approach of Eq.~(\ref{eq:ampF}). Fitting the available data leads to $\mathcal{F}\approx 0.5$. One notices that the angular dependence of the data is well reproduced.
We find that the Regge model including $N$ and $\Delta$-trajectory exchange describes the available data very well. The results give a better description compared to the case when only $N$-trajectories are taken into account. In particular for larger values of $s$, the results of the $N$-trajectory lie below the data. The cross section including the reduction factor $\mathcal{F}$ of the $\Delta$-pole residue is in very good agreement with the experiment, especially in the region $s \gg |t|,|u|$.

Additionally, we considered a $\pi NN$ coupling of the pseudovector type, but we did not find a good description of the data in the forward and backward regions for this coupling.

\begin{figure}
 \includegraphics[width=8cm]{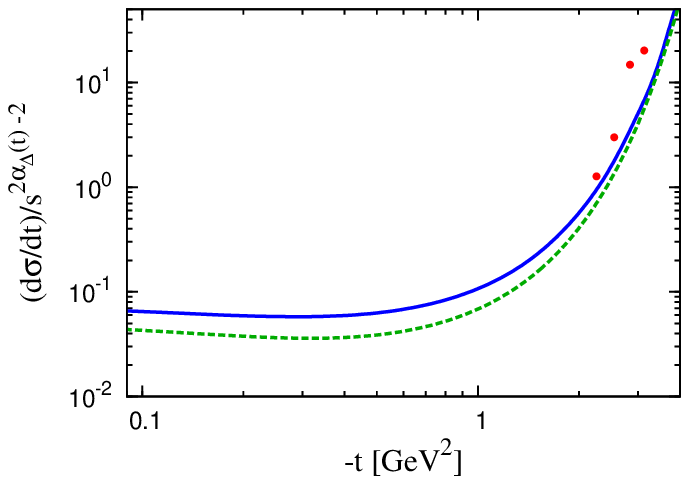}
\caption{Cross section $d\sigma/dt$ divided by $s^{2\alpha_\Delta(t) -2}$ for the process $\bar{p}p\to \pi^0 \gamma$ as a function of $-t$ for different values of $\sqrt{s}$ using the description according to Eq.~(\ref{eq:ampF}); blue solid curve: $\sqrt{s}=3.686$~GeV; green dashed curve: $\sqrt{s}=10$~GeV; the data correspond to the cross section measured at $\sqrt{s}=3.686$~GeV \cite{Armstrong:1997gv}.  \label{fi:dt}}
\end{figure}

In order to test the applicability of the model, we present in Fig.~\ref{fi:dt} the cross section $d\sigma/dt$, divided by the expected $s$-dependence of the leading Regge trajectory, $s^{2\alpha_\Delta(t) -2}$, as a function of $-t$ for two different c.m. energies. One notices, that for $-t \rightarrow 0$ $(d\sigma/dt)/s^{2\alpha_\Delta(t) -2}$ shows a behavior, which is approximately independent of $s$, as expected from Regge theory, and approaches a constant value. The existing cross section data, as indicated by the data taken at $\sqrt{s}=3.686$~GeV in Fig.~\ref{fi:dt}, have not yet reached the region of such small values of $-t$, where an extrapolation of $(d\sigma/dt)/s^{2\alpha_\Delta(t) -2}$ could be performed by a constant.

\section{The $\bar{p}p \rightarrow \pi^0e^+e^-$ Process within a Regge Framework}
\label{sec:pi0ll}
After specifying the Regge pole model, we study the process
\begin{equation}
 \bar{p}(p_1) + p(p_2) \rightarrow \pi^0(q_\pi) + \gamma^*(q)  \rightarrow \pi^0(q_\pi) + e^-(k_1) + e^+(k_2), 
\end{equation}
and will start with a model independent analysis of the annihilation cross section. The cross section for the process is defined as
\begin{align}
 d\sigma =& \frac{1}{4\sqrt{(p_1\cdot p_2) ^2-m_N^4}}\left(\frac{d^3\vec{q}_\pi}{(2\pi)^3 2 E_\pi}\right)\left(\frac{d^3\vec{k}_1}{(2\pi)^3 2 k_1^0}\right)\left(\frac{d^3\vec{k}_2}{(2\pi)^3 2 k_2^0}\right) \nonumber \\[1mm]
 & \times \,  (2\pi)^4\delta(p_1+p_2-q_\pi-k_1-k_2)|\mathcal{T}|^2,
\end{align}
where the amplitude $\mathcal{T}$ can be decomposed into a hadronic and a leptonic contribution:
\begin{align}
|\mathcal{T}|^2 =& \sum_{\lambda_\gamma=0,\pm 1}\ \frac{1}{4}\sum_{\mathrm{spins}} \Big| \Big(\mathcal{M}^\mu_{\pi^0\gamma^*} \cdot \varepsilon^*_\mu(q,\lambda_\gamma)\Big) \,\frac{1}{q^2}\, \Big(\varepsilon_\nu(q,\lambda_\gamma) \bar{u}(k_1)\, e \gamma^\nu \,v(k_2)\Big)\Big|^2.
\label{eq:ampT}
\end{align}
$\mathcal{M}^\mu_{\pi^0\gamma^*}$ is the amplitude of the hadronic process $\bar{p}p\rightarrow \pi^0 \gamma^*$.
Both contributions of Eq. (\ref{eq:ampT}),
\begin{equation}
  \big| \mathcal{M}^\mu_{\pi^0\gamma^*} \cdot \varepsilon^*_\mu(q,\lambda_\gamma)\big|^2 \quad \mbox{and} \quad \big|\varepsilon_\nu(q,\lambda_\gamma) \bar{u}(k_1)\, e \gamma^\nu \,v(k_2)\big|^2,
\end{equation}
are invariant, thus we can choose any reference frame for our calculation. The advantage of such a separation is that one can easily calculate the hadronic and leptonic processes in two different reference frames.
\begin{figure}
\includegraphics[width=8.0cm]{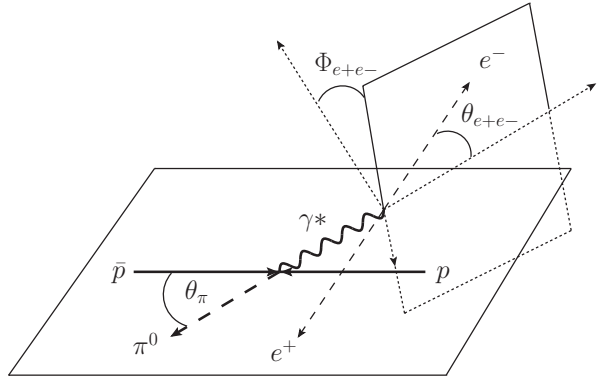}
\caption{Kinematics of the $\bar{p}p \rightarrow \pi^0e^+e^-$ process. \label{fi:kin}} 
\end{figure}

We consider the hadronic process in the c.m.-frame of the nucleon pair, where the momenta of the incoming proton and antiproton have opposite directions, illustrated in Fig.~\ref{fi:kin}.
The leptonic subprocess $\gamma^* \rightarrow e^+ e^-$ is computed in the $\gamma^*$-rest frame, with the 4-momentum of the virtual photon given by
\begin{equation}
 q=(\sqrt{q^2},0,0,0)
\end{equation}
and where the lepton momenta can be written as
\begin{align}
\begin{split}
 k_1 =& \frac{\sqrt{q^2}}{2}\,\big(1,\ \sin\theta_{e^+e^-}\cos\Phi_{e^+e^-},\ \sin\theta_{e^+e^-} \sin\Phi_{e^+e^-},\cos\theta_{e^+e^-} \big), \\
 k_2 =& \frac{\sqrt{q^2}}{2}\,\big(1,- \sin\theta_{e^+e^-}\cos\Phi_{e^+e^-},-\sin\theta_{e^+e^-} \sin\Phi_{e^+e^-},- \cos\theta_{e^+e^-} \big).
\end{split}
\end{align}
Therefore, we choose the angles $\theta_{e^+e^-}$ and $\Phi_{e^+e^-}$ as two independent kinematic variables describing the leptonic subprocess. The hadronic part of the amplitude depends on the c.m. energy $\sqrt{s}$, the virtuality of the photon $q^2$, and the Mandelstam variable $t$, which can be related to the pion-scattering angle $\theta_\pi$.

The differential cross section of the reaction is given by
\begin{align}
  \frac{d\sigma}{dt\,dq^2\,d\Omega_{e^+e^-}} = \frac{1}{16\pi^2s(s-4m_N^2)}\,\frac{e^2}{(4\pi)^2\,8\,q^2}\,\frac{4\pi}{3}\cdot \mathcal{W}(\theta_{e^+e^-}, \Phi_{e^+e^-}),
\label{eq:csgeneral}
\end{align}
with the leptonic solid angle $d\Omega_{e^+e^-}$. The $e^+e^-$ decay angular distribution $\mathcal{W}$ can be expressed as
\begin{align}
\mathcal{W}(\theta_{e^+e^-}, \Phi_{e^+e^-}) =& \frac{3}{4\pi}\Big[ \sin^2\theta_{e^+e^-} \rho_{00} + (1+\cos^2\theta_{e^+e^-})\rho_{11} \nonumber \\[1mm]
 & \ +\ \sqrt{2}\sin 2\theta_{e^+e^-}\cos\Phi_{e^+e^-}\mathrm{Re}[\rho_{10}] \ + \ \sin^2\theta_{e^+e^-}\cos2\Phi_{e^+e^-}\mathrm{Re}[\rho_{1-1}] \ \Big],
\label{eq:angularD}
\end{align}
where the density matrix $\rho$ is given by
\begin{equation}
 \rho_{\lambda\lambda^\prime} = \left(\mathcal{M}_{\pi^0\gamma^*}^\mu \varepsilon^*_\mu(\lambda_\gamma) \right)\left(\mathcal{M}_{\pi^0\gamma^*}^\mu \varepsilon^*_\mu(\lambda^\prime_\gamma) \right)^*, \quad \text{for} \ \lambda_\gamma,\lambda_\gamma^\prime=0,\pm 1.
\end{equation}

The expression presented in Eq.~(\ref{eq:csgeneral}) is model independent. However, in order to obtain numerical results we have to use a model to characterize the hadronic subprocess. We choose the Regge pole model which has been introduced in Section~\ref{sec:pi0gamma}. Since a virtual photon is produced, we have to specify the electromagnetic form factors parametrizing the $\gamma^* NN$-vertices and $\gamma^* N \Delta$-vertices. For the electromagnetic form factors we consider a vector meson dominance (VMD) model, which assumes that the electromagnetic interaction is described through the exchange of the lowest lying vector mesons $\rho$, $\omega$, and $\phi$. In \cite{Iachello:2004aq} a parametrization for both form factors $F_1$ and $F_2$ in the spacelike as well as in the timelike region is presented, which we will use for the purpose of the computation of the cross section. Eventually, the aim of the present work is to provide a further constraint on future extractions of timelike nucleon form factors.

For the magnetic dipole form factor of the $N\rightarrow \Delta$ transition, we use the results in the large $N_c$ limit, which can be written as \cite{Pascalutsa:2007wz}:
\begin{equation}
 g_M(q^2) = \frac{g_M(0)}{\kappa_V}[F_2^p(q^2)-F_2^n(q^2)],
\end{equation}
where $F_2^p$ ($F_2^n$) is the Pauli form factors of the proton (neutron), for which we use the description of the VMD model, and $\kappa_V= \kappa_p - \kappa_n = 3.70$.

\begin{figure*}
\includegraphics[width=7.8cm]{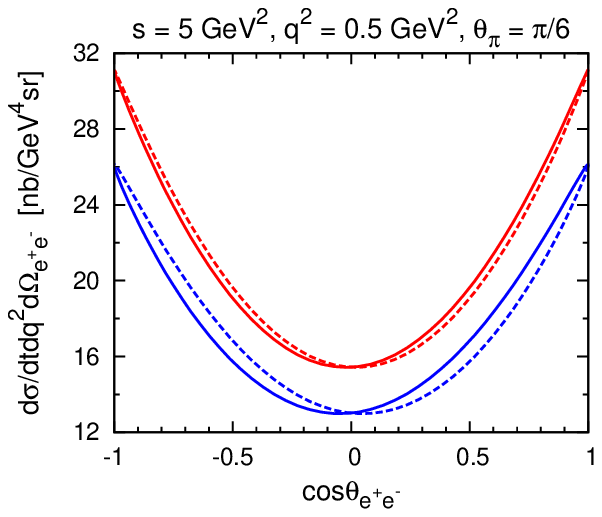}
\includegraphics[width=7.8cm]{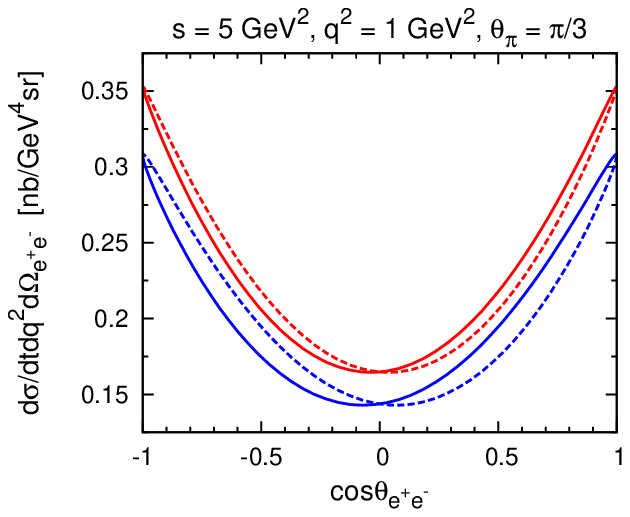}\\
\includegraphics[width=7.8cm]{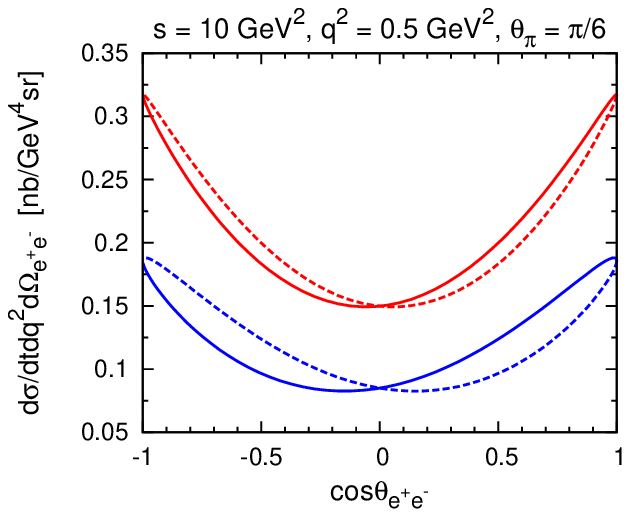}
\includegraphics[width=7.8cm]{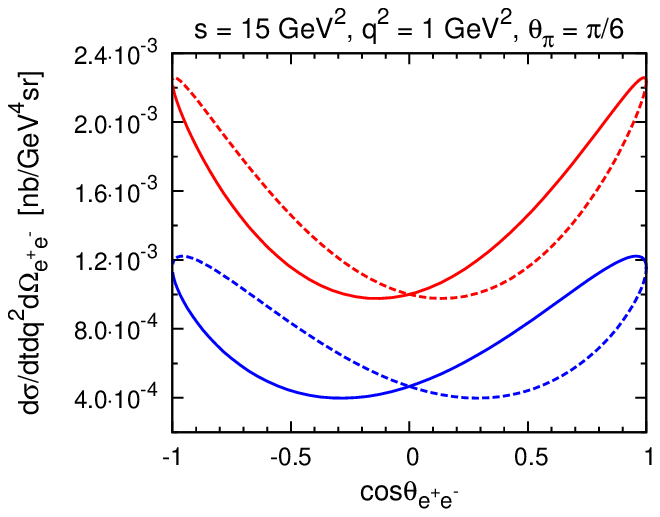}
\caption{Differential cross section $d\sigma/dt\, dq^2\, d\Omega_{e^+e^-}$ of $\bar{p}p \rightarrow \pi^0e^+e^-$ as a function of $\cos\theta_{e^+e^-}$. Red curves correspond to the $N$-trajectory contribution; red solid curve: $\Phi_{e^+e^-} = 0$; red dashed curve: $\Phi_{e^+e^-} = \pi$. Blue curves correspond to the $(N+\Delta)$-trajectory contribution including a reduction of the $\Delta$ pole residue ($\mathcal{F}\approx 0.5$) according to Eq.~(\ref{eq:ampF}); blue solid curve: $\Phi_{e^+e^-} = 0$; blue dashed curve: $\Phi_{e^+e^-} = \pi$. \label{fi:cs_pi0ll}}
\end{figure*}

The results of the differential cross section $d\sigma/dt\, dq^2\, d\Omega_{e^+e^-}$ as a function of $\cos\theta_{e^+e^-}$ are presented in Fig.~\ref{fi:cs_pi0ll} for several kinematical conditions. We display the $N$-trajectory and $(N+\Delta)$-trajectory contributions as introduced in Eq.~(\ref{eq:ampF}) for the angles $\Phi_{e^+e^-}=0$ and $\Phi_{e^+e^-}=\pi$.
The dependence on the angle $\Phi_{e^+e^-}$ appears as an asymmetric behavior of the cross section with respect to $\cos\theta_{e^+e^-}$. For $\Phi_{e^+e^-} = \pi/2$ the resulting cross section is symmetric, which can be derived from the general form of the decay angular distribution $\mathcal{W}$, given by Eq.~(\ref{eq:angularD}).

Using the Born model suggested in \cite{Adamuscin:2007iv}, one obtains a cross section, which is 1 to 4 orders of magnitudes larger than the results of the Regge pole model, depending on the variation of the kinematic parameters $s$, $q^2$ and $\theta_\pi$.

\begin{figure}
 \includegraphics[width=7.8cm]{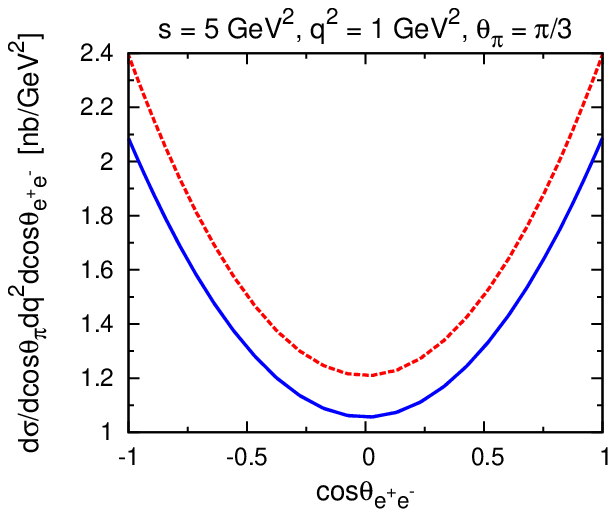}
 \includegraphics[width=7.8cm]{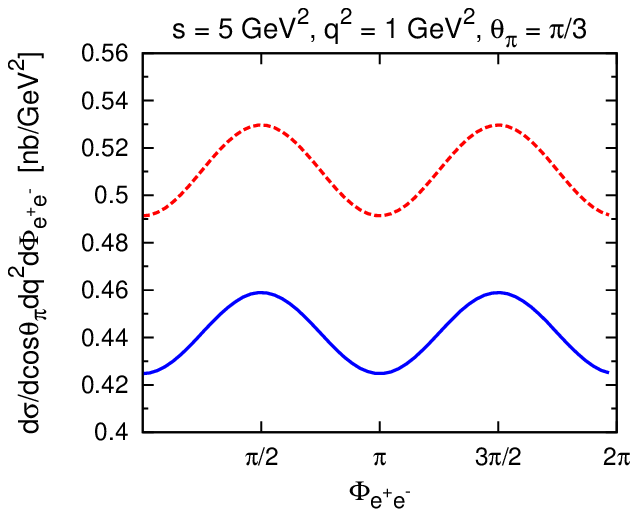}
\caption{Cross sections defined by Eq.~(\ref{eq:intcs_phi}) (left) and Eq.~(\ref{eq:intcs_theta}) (right): Red (dashed) curve: $N$-trajectory contribution; blue (solid) $(N+\Delta)$-trajectory contribution according to Eq.~(\ref{eq:ampF}). \label{fi:intcs}}
\end{figure}

The integrated cross section can be used to access the density matrix $\rho$. The cross section integrated over the azimuthal angle $\Phi_{e^+e^-}$
\begin{equation}
 \frac{d\sigma}{d\cos\theta_\pi\, dq^2\, d\cos\theta_{e^+e^-}} = \int_0^{2\pi}d\Phi_{e^+e^-}\frac{d\sigma}{d\cos\theta_\pi\,dq^2\, d\Omega_{e^+e^-}},
\label{eq:intcs_phi}
\end{equation}
is sensitive to $\rho_{00}$ and $\rho_{11}$ and the cross section integrated over the polar angle $\theta_{e^+e^-}$
\begin{equation}
  \frac{d\sigma}{d\cos\theta_\pi\, dq^2\, d\Phi_{e^+e^-}} = \int_{-1}^{1}d\cos\theta_{e^+e^-}\frac{d\sigma}{d\cos\theta_\pi\, dq^2 \,d\Omega_{e^+e^-}},
\label{eq:intcs_theta}
\end{equation}
can be analyzed in order to extract $\rho_{1-1}$ from the $\Phi$ dependence of the cross section. As selective predictions, we show the results of the cross sections (\ref{eq:intcs_phi}), and (\ref{eq:intcs_theta}) in Fig.~\ref{fi:intcs} for $s=~5$ GeV$^2$, $q^2=1$ GeV$^2$ and $\theta_\pi = \pi/3$ using both $N$-tajectory and $(N+\Delta)$-trajectory exchange, given by Eq.~(\ref{eq:ampF}).

\begin{figure}
 \includegraphics[width=7.8cm]{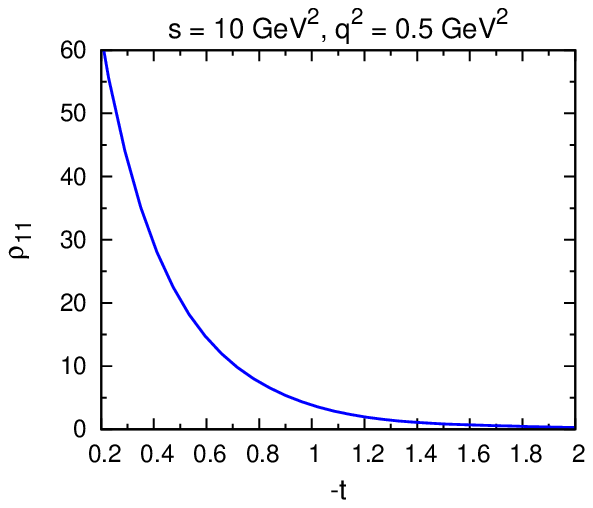}
 \includegraphics[width=7.8cm]{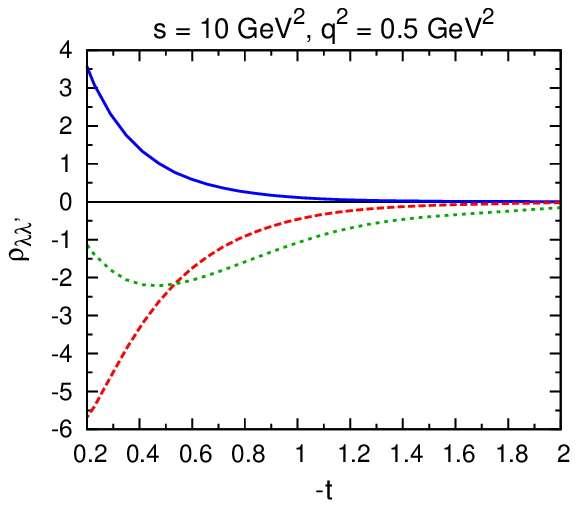}
\caption{Density matrices as a function of $-t$ (in GeV$^2$) for $s=10$ GeV$^2$ and $q^2$=0.5 GeV$^2$ using a Regge model with $N$ and $\Delta$ trajectory exchange according to Eq.~(\ref{eq:ampF}). Left panel: density matrix $\rho_{11}$; Right panel: $\rho_{00}$ (blue solid curve), $\rho_{10}$ (red dashed curve), $\rho_{1-1}$ (green dotted curve). \label{fi:dm}}
\end{figure}

The $t$-dependence of the density matrices $\rho_{\lambda\lambda^\prime}$ is presented in Fig.~\ref{fi:dm} using a $(N+\Delta)$-trajectory exchange as introduced in Eq.~(\ref{eq:ampF}) for the region $s\gg|t|$, which is dominated by the $t$-channel Regge amplitude. The density matrix $\rho_{11}$, shown in the left panel of Fig.~\ref{fi:dm}, yields the dominant contribution to the cross section, since it is about 1 order of magnitude larger compared to the three other structures, presented in the right panel of Fig.~\ref{fi:dm}.


\section{Conclusions}
\label{conclusions}
In this paper we studied the process $p\bar{p} \to \pi^0 e^+ e^-$, giving a model independent expression of the cross section in terms of the lepton pair angular distribution and presenting results within a Regge pole approach. Such a model description is applicable in the forward and backward angular ranges.
We found that a model based on nucleon and $\Delta$-Regge trajectory exchange provides a good description of the data for the process of real photoproduction $p\bar{p} \to \pi^0 \gamma$ in the energy range of $s\simeq 8.5 - 14$ GeV$^2$ in the regions $s\gg|t|,|u|$. Applying this model to $p\bar{p} \to \pi^0 e^+ e^-$ allowed us to predict the angular dependence of the differential cross section, which can be used to extract the timelike form factors in the unphysical region as well as their phases, in kinematics which will be accessible by the PANDA@FAIR experiment.



\section*{Acknowledgments}
This work was supported in part by the Research Centre ``Elementarkraefte
und Mathematische Grundlagen" at the Johannes Gutenberg University
Mainz and in part by the Helmholtz Institute Mainz (HIM). 
The authors like to thank F. Maas, M. Zambrana and C. Adamu\v{s}\v{c}\'in for useful discussions.


\begin{thebibliography}{99}

\bibitem{HydeWright:2004gh}
 C.~E.~Hyde-Wright and K.~de Jager,
 Ann.\ Rev.\ Nucl.\ Part.\ Sci.\  {\bf 54}, 217 (2004). 

\bibitem{Arrington:2006zm}
 J.~Arrington, C.~D.~Roberts and J.~M.~Zanotti,
 J.\ Phys.\ G {\bf 34}, S23 (2007). 

\bibitem{Perdrisat:2006hj}
 C.~F.~Perdrisat, V.~Punjabi and M.~Vanderhaeghen,
 Prog.\ Part.\ Nucl.\ Phys.\  {\bf 59}, 694 (2007). 

\bibitem{Aubert:2005cb} 
  B.~Aubert {\it et al.}  [BABAR Collaboration],
  Phys.\ Rev.\ D {\bf 73}, 012005 (2006).

\bibitem{Bardin:1994am} 
  G.~Bardin, G.~Burgun, R.~Calabrese, G.~Capon, R.~Carlin, P.~Dalpiaz, P.~F.~Dalpiaz and J.~Derre {\it et al.},
  Nucl.\ Phys.\ B {\bf 411}, 3 (1994).

\bibitem{Lutz:2009ff}
  M.~F.~M.~Lutz {\it et al.}  [PANDA Collaboration],
  arXiv:0903.3905 [hep-ex].

\bibitem{Dubnickova:1995ns} 
  A.~Z.~Dubnickova, S.~Dubnicka and M.~P.~Rekalo,
  Z.\ Phys.\ C {\bf 70}, 473 (1996).

\bibitem{Adamuscin:2007iv} 
  C.~Adamuscin, E.~A.~Kuraev, E.~Tomasi-Gustafsson and F.~E.~Maas,
  Phys.\ Rev.\ C {\bf 75}, 045205 (2007).

\bibitem{Boucher2011} 
  J.~Boucher,
  Ph.D. thesis, Universite Paris Sud XI, Orsay, and Johannes Gutenberg University, Mainz (2011).

\bibitem{Gakh:2012uk} 
  G.~I.~Gakh, E.~Tomasi-Gustafsson, A.~Dbeyssi and A.~G.~Gakh,
  Phys.\ Rev.\ C {\bf 86}, 025204 (2012).

\bibitem{Lansberg:2007se} 
  J.~P.~Lansberg, B.~Pire and L.~Szymanowski,
  Phys.\ Rev.\ D {\bf 76}, 111502 (2007).

\bibitem{Lansberg:2012ha} 
  J.~P.~Lansberg, B.~Pire, K.~Semenov-Tian-Shansky and L.~Szymanowski,
  Phys.\ Rev.\ D {\bf 86}, 114033 (2012)

\bibitem{Collins:1977jy} 
  P.~D.~B.~Collins,
  Cambridge 1977, 445p.

\bibitem{Storrow:1986zw} 
  J.~K.~Storrow,
  Rept.\ Prog.\ Phys.\  {\bf 50}, 1229 (1987).

\bibitem{Guidal:1997hy} 
  M.~Guidal, J.~M.~Laget and M.~Vanderhaeghen,
  Nucl.\ Phys.\ A {\bf 627}, 645 (1997).

\bibitem{Vanderhaeghen:1997ts}
  M.~Vanderhaeghen, M.~Guidal and J.~M.~Laget,
  Phys.\ Rev.\ C {\bf 57} (1998) 1454.

\bibitem{Blok:2008jy} 
  H.~P.~Blok {\it et al.}  [Jefferson Lab Collaboration],
  Phys.\ Rev.\ C {\bf 78}, 045202 (2008).

\bibitem{Coman:2009jk} 
  M.~Coman {\it et al.}  [Jefferson Lab Hall A Collaboration],
  Phys.\ Rev.\ C {\bf 81}, 052201 (2010).

\bibitem{Armstrong:1997gv} 
  T.~A.~Armstrong {\it et al.}  [Fermilab E760 Collaboration],
  Phys.\ Rev.\ D {\bf 56}, 2509 (1997).

\bibitem{Kroll:2005ni} 
  P.~Kroll and A.~Schafer,
  Eur.\ Phys.\ J.\ A {\bf 26}, 89 (2005).

\bibitem{Tiator:2011pw} 
  L.~Tiator, D.~Drechsel, S.~S.~Kamalov and M.~Vanderhaeghen,
  Eur.\ Phys.\ J.\ ST {\bf 198}, 141 (2011).

\bibitem{Iachello:2004aq}
 F.~Iachello and Q.~Wan,
 Phys.\ Rev.\  C {\bf 69}, 055204 (2004).

\bibitem{Pascalutsa:2007wz} 
  V.~Pascalutsa and M.~Vanderhaeghen,
  Phys.\ Rev.\ D {\bf 76}, 111501 (2007).


\end{thebibliography}
\end{document}